\documentclass[preprintnumbers,amsmath,amssymbm,prd]{revtex4}
\usepackage{epsfig}
\usepackage{graphicx}
\usepackage{amssymb}

\begin{document}
\title{No hair for spherically symmetric neutral reflecting stars: nonminimally coupled massive scalar fields}
%\title{No nonminimally coupled massive scalar hair for spherically symmetric neutral reflecting stars}
\author{Shahar Hod}
\affiliation{The Ruppin Academic Center, Emeq Hefer 40250, Israel}
\affiliation{ }
\affiliation{The Hadassah Institute, Jerusalem 91010, Israel}
\date{\today}

\begin{abstract}
\ \ \ Recent no-hair theorems have revealed the intriguing fact that
horizonless stars with compact reflecting surfaces {\it cannot}
support non-linear matter configurations made of scalar, vector, and
tensor fields. In the present paper we extend the regime of validity
of these no-hair theorems by explicitly proving that spherically
symmetric compact reflecting stars cannot support static
configurations made of {\it massive} scalar fields with {\it
non}-minimal coupling to gravity. Interestingly, our no-hair theorem
is valid for {\it generic} values of the dimensionless
field-curvature coupling parameter $\xi$.
\end{abstract}
\bigskip
\maketitle

\section{Introduction}

The influential no-scalar-hair theorems of Bekenstein \cite{Bek1}
and Teitelboim \cite{Teit} (see also
\cite{Bekto,Chas,Heu,Bek2,BekMay,Bek20} and references therein) have
revealed the interesting fact that asymptotically flat classical
black-hole spacetimes, which are characterized by absorbing one-way
membranes (event horizons), cannot support static non-linear matter
configurations (external `hair') made of spatially regular massive
scalar fields with minimal coupling to gravity
\cite{Noteev,Noteit,Hodrc,Herkr}.

The rigorous derivation of a no-hair theorem for scalar matter
configurations {\it nonminimally} coupled to gravity turns out to be
a mathematically challenging task. In particular, the interesting
no-hair theorems presented by Bekenstein and Mayo
\cite{BekMay,Bek20} can be used to rule out the existence of
spherically symmetric asymptotically flat black-hole spacetimes
supporting neutral nonminimally coupled massive scalar fields in the
restricted physical regimes $\xi<0$ and $\xi\geq1/2$
\cite{Notewx,Notenv,NoteBB,BB,BekenBB,Notesm}.

Recently, we have raised \cite{Hodns} the following physically
intriguing question: Can the canonical no-scalar-hair theorems
presented in \cite{Bek1,Teit,Bekto,Chas,Heu,Bek2,BekMay,Bek20} for
classical black-hole spacetimes with absorbing event horizons be
extended to the physical regime of asymptotically flat {\it
horizonless} curved spacetimes?

It is worth mentioning that this physically important question is
also motivated by the recent scientific interest in the physical and
mathematical properties of regular spacetimes describing highly
compact horizonless objects (see \cite{MPF,BCP,Hodrj,Hodrb} and
references therein) which have been proposed as possible exotic
quantum-gravity alternatives to the canonical classical black-hole
spacetimes.

Motivated by the above stated physically interesting question, in
Ref. \cite{Hodns} we have explored the physical implications of
replacing the standard ingoing ({\it absorbing}) boundary condition
which characterizes the behavior of matter fields at the classical
horizon of a black-hole spacetime \cite{Bek1,Bek20} by a reflecting
({\it repulsive}) boundary condition at the surface of a spherically
symmetric horizonless compact star. Intriguingly, the analysis
presented in \cite{Hodns} has revealed the fact that horizonless
compact stars with reflecting surfaces \cite{Noteref} share the
no-scalar-hair property with the more familiar classical black-hole
spacetimes with absorbing one-way membranes (event horizons). In
particular, it has been explicitly proved \cite{Hodns} that
spherically symmetric compact objects with reflecting boundary
conditions cannot support spatially regular matter configurations
made of non-linear scalar (spin-0) fields with {\it minimal}
coupling to gravity. Moreover, in a very interesting paper,
Bhattacharjee and Sarkar \cite{Bha} have extended the results of
\cite{Hodns} and proved that horizonless stars with reflecting
boundary conditions cannot support vector (spin-1) and tensor
(spin-2) fields. Recently, we have explicitly proved that neutral
reflecting stars \cite{Noteref} cannot support static spherically
symmetric configurations of {\it nonminimally} coupled {\it
massless} scalar fields \cite{Hodpd}.

It is physically important to explore the regime of validity of the
intriguing no-hair behavior recently observed for {\it horizonless}
regular spacetimes describing compact reflecting stars
\cite{Hodns,Bha,Hodpd}. The main goal of the present paper is to
analyze the physical and mathematical properties of the non-linearly
coupled Einstein-scalar field equations for {\it massive} scalar
fields with {\it nonminimal} ($\xi\neq0$) coupling to gravity. In
particular, below we shall present a novel no-hair theorem which
explicitly proves that static spherically symmetric neutral stars
with reflecting (that is, {\it repulsive} rather than {\it
absorbing}) boundary conditions cannot support spatially regular
matter configurations made of non-linear {\it massive} scalar fields
with {\it generic} values of the dimensionless nonminimal coupling
parameter $\xi$.

\section{Description of the system}

We shall analyze the static sector of a physical system which is
composed of a spherically symmetric neutral reflecting object (a
compact reflecting star \cite{Noteref}) of radius $R_{\text{s}}$
which is non-linearly coupled to a massive scalar field. The scalar
field is assumed to be non-minimally coupled to gravity. The curved
line element of the static and spherically symmetric spacetime can
be expressed in the form \cite{BekMay,Notecor,Noteunit}
\begin{equation}\label{Eq1}
ds^2=-e^{\nu}dt^2+e^{\lambda}dr^2+r^2(d\theta^2+\sin^2\theta
d\phi^2)\ ,
\end{equation}
where $\nu=\nu(r)$ and $\lambda=\lambda(r)$. Asymptotic flatness of
the spacetime is characterized by the asymptotic large-$r$ behaviors
\cite{BekMay}
\begin{equation}\label{Eq2}
\nu\sim M/r\ \ \ \text{and}\ \ \ \lambda\sim M/r\ \ \ \ \text{for}\
\ \ \ r\to\infty\
\end{equation}
of the metric functions, where $M$ is the total ADM mass (as
measured by asymptotic observers) of the composed static star-field
configurations.

The action of the non-minimally coupled massive scalar field $\psi$
in the curved spacetime of the spherically symmetric reflecting star
is given by \cite{BekMay,NoteSEH}
\begin{equation}\label{Eq3}
S=S_{EH}-{1\over2}\int\big(\partial_{\alpha}\psi\partial^{\alpha}\psi+\mu^2\psi^2+\xi
R\psi^2\big)\sqrt{-g}d^4x\ ,
\end{equation}
where $\mu$ is the proper mass of the field \cite{Notemu} and $R(r)$
is the scalar curvature of the spacetime. An asymptotically flat
spacetime is characterized by the large-$r$ asymptotic behavior
\cite{BekMay}
\begin{equation}\label{Eq4}
R(r\to\infty)\to 0\  .
\end{equation}
The dimensionless physical parameter $\xi$ in the action (\ref{Eq3})
quantifies the strength of the nonminimal coupling of the massive
scalar field to the spacetime curvature.

The action (\ref{Eq3}) yields the characteristic differential
equation \cite{BekMay}
\begin{equation}\label{Eq5}
\partial_{\alpha}\partial^{\alpha}\psi-(\mu^2+\xi R)\psi=0\
\end{equation}
for the nonminimally coupled massive scalar field. Substituting into
(\ref{Eq5}) the metric components of the spherically symmetric
curved spacetime (\ref{Eq1}), one finds that the spatial behavior of
the static nonminimally coupled massive scalar field is governed by
the non-linear radial differential equation \cite{Notetag}
\begin{equation}\label{Eq6}
\psi{''}+{1\over2}\big({{4}\over{r}}+\nu{'}-\lambda{'}\big)\psi{'}-(\mu^2+\xi
R) e^{\lambda}\psi=0\  .
\end{equation}
The nonminimally coupled massive scalar field $\psi$ is assumed to
vanish on the surface $r=R_{\text{s}}$ of the central spherically
symmetric compact {\it reflecting} star \cite{Noterefbs,PressTeu2}:
\begin{equation}\label{Eq7}
\psi(r=R_{\text{s}})=0\  .
\end{equation}
In addition, as discussed in \cite{BekMay}, a curved spacetime
describing a physically acceptable system must be characterized by
an asymptotic finite and positive value of the effective
gravitational constant $G_{\text{eff}}=G(1-8\pi G\xi\psi^2)$
\cite{BekMay}. This physical requirement enforces the asymptotic
spatial bounds
\begin{equation}\label{Eq8}
-\infty<8\pi\xi\psi^2<1\ \ \ \ \text{for}\ \ \ \ r\to\infty\
\end{equation}
on the characteristic radial eigenfunctions of the nonminimally
coupled massive scalar fields.

From the action (\ref{Eq3}) one also finds the following expressions
\cite{BekMay}
\begin{equation}\label{Eq9}
T^{t}_{t}=e^{-\lambda}{{\xi(4/r-\lambda{'})\psi\psi{'}+(2\xi-1/2)(\psi{'})^2+2\xi\psi\psi{''}}
\over{1-8\pi\xi\psi^2}}-{{\mu^2\psi^2}\over{2(1-8\pi\xi\psi^2)}}\ ,
\end{equation}
\begin{equation}\label{Eq10}
T^{t}_{t}-T^{r}_{r}=e^{-\lambda}{{(2\xi-1)(\psi{'})^2+2\xi\psi\psi{''}-\xi(\nu+\lambda){'}\psi\psi{'}}
\over{1-8\pi\xi\psi^2}}\  ,
\end{equation}
and
\begin{equation}\label{Eq11}
T^{t}_{t}-T^{\phi}_{\phi}=e^{-\lambda}{{\xi(2/r-\nu{'})\psi\psi{'}}
\over{1-8\pi\xi\psi^2}}
\end{equation}
for the components of the energy-momentum tensor which characterizes
the nonminimally coupled massive scalar field in the static curved
spacetime (\ref{Eq1}) of the star.
%As explicitly proved in \cite{BekMay},
%physically acceptable regular spacetimes are characterized by finite
%values of the mixed components of the energy-momentum tensor:
%\begin{equation}\label{Eq11}
%\{|T^{t}_{t}|,|T^{r}_{r}|,|T^{\theta}_{\theta}|,|T^{\phi}_{\phi}|\}<\infty\
%.
%\end{equation}
%In particular, the energy density $\rho\equiv -T^{t}_{t}$ of an
%Below we shall make use of the fact that the energy density
%$\rho\equiv -T^{t}_{t}$ of an asymptotically flat spacetime with a
%finite ADM mass must decay asymptotically faster than $1/r^3$
%\cite{Hodpb}:
%\begin{equation}\label{Eq11}
%r^3\rho(r)\to0\ \ \ \text{for}\ \ \ r\to\infty\  .
%\end{equation}

In addition, as explicitly proved by Bekenstein and Mayo
\cite{BekMay}, causality requirements imply that the components of
the energy-momentum tensors of physically acceptable systems must
respect the following inequalities:
\begin{equation}\label{Eq12}
|T^{\theta}_{\theta}|=|T^{\phi}_{\phi}|\leq|T^{t}_{t}|\geq|T^{r}_{r}|\
.
\end{equation}
For later purposes, we note that the simple relations \cite{BekMay}
\begin{equation}\label{Eq13}
\text{sgn}(T^{t}_{t})=\text{sgn}(T^{t}_{t}-T^{r}_{r})=\text{sgn}(T^{t}_{t}-T^{\phi}_{\phi})\
\end{equation}
provide necessary conditions for the validity of the energy
conditions (\ref{Eq12}) which, as proved in \cite{BekMay},
characterize the energy-momentum components of physically acceptable
systems.

\section{The no-hair theorem for the composed reflecting-stars-nonminimally-coupled-massive-scalar
fields configurations}

In the present section we shall explicitly prove that spatially
regular static matter configurations made of nonminimally coupled
massive scalar fields with the action (\ref{Eq3}) {\it cannot} be
supported by spherically symmetric {\it horizonless} reflecting
stars.

We shall first prove that, for physically acceptable systems, the
characteristic radial eigenfunction $\psi$ of the massive scalar
fields must approach zero at spatial infinity. Taking cognizance of
the asymptotic functional relations (\ref{Eq2}) and (\ref{Eq4}),
which characterize asymptotically flat spacetimes with finite ADM
masses, one finds that, in the asymptotic far region $M/r\ll 1$, the
radial equation (\ref{Eq6}) of the nonminimally coupled massive
scalar fields can be approximated by
\begin{equation}\label{Eq14}
\psi{''}+{{2}\over{r}}\psi{'}-\mu^2\psi=0\ .
\end{equation}
The general mathematical solution of the asymptotic radial equation
(\ref{Eq14}) is given by
\begin{equation}\label{Eq15}
\psi(r)=A\cdot(\mu r)^{-1}e^{-\mu r}+B\cdot(\mu r)^{-1}e^{\mu r}\ \
\ \ \text{for}\ \ \ \ r\gg M\ ,
\end{equation}
where $\{A,B\}$ are dimensionless constants.
%Substituting the radial
%solution (\ref{Eq15}) with $B\neq0$ into (\ref{Eq8}) and using the
%asymptotic far region relation (\ref{Eq2}), one finds the
%non-vanishing asymptotic expression \cite{Notex4}
%\begin{equation}\label{Eq16}
%\rho={{4\xi-1}\over{8\pi\xi}}\mu^2\cdot[1+O(M/r)]\
%\end{equation}
%for the energy density of the nonminimally coupled massive scalar
%fields. One immediately realizes that (\ref{Eq16}) violates the
%characteristic relation (\ref{Eq11}) for asymptotically flat
%spacetimes \cite{Hodpb}.
%We therefore find that a physically acceptable (finite mass)
%non-linear massive scalar field configuration is characterized by
%the asymptotic radial eigenfunction [see Eq. (\ref{Eq15}) with
%$B=0$] \cite{Notebo}

One immediately realizes that the asymptotic radial solution
(\ref{Eq15}) with $B\neq0$ violates the upper bound (\ref{Eq8})
which, as discussed in \cite{BekMay}, characterizes physically
acceptable systems. We therefore find that spatially regular massive
scalar field configurations supported in physically acceptable
asymptotically flat spacetimes [that is, spacetimes which are
characterized by finite and positive asymptotic values of the
effective gravitational constant $G_{\text{eff}}=G(1-8\pi
G\xi\psi^2)$ \cite{BekMay}] are characterized by the asymptotic
radial eigenfunction [see Eq. (\ref{Eq15}) with $B=0$]
%\cite{Notebo}
\begin{equation}\label{Eq16}
\psi(r)=A\cdot(\mu r)^{-1}e^{-\mu r}\ \ \ \ \text{for}\ \ \ \ r\gg
M\ .
\end{equation}

For later purposes, we note that the characteristic asymptotic
behavior (\ref{Eq16}) of the massive scalar fields together with the
inner boundary condition (\ref{Eq7}) at the surface of the
horizonless spherically symmetric compact reflecting star imply that
the radial scalar eigenfunction $\psi$ must have (at least) one
extremum point, $r=r_{\text{peak}}$, which is located in the
interval $r_{\text{peak}}\in(R_{\text{s}},\infty)$. At this extremum
point the radial eigenfunction of the static nonminimally coupled
massive scalar fields is characterized by the simple relations
\begin{equation}\label{Eq17}
\{\psi\neq0\ \ \ ; \ \ \ \psi{'}=0\ \ \ ; \ \ \
\psi\cdot\psi{''}<0\}\ \ \ \ \text{for}\ \ \ \ r=r_{\text{peak}}\  .
\end{equation}

\subsection{The no-massive-scalar-hair theorem for generic inner boundary conditions in
the physical regimes $\xi<0$ and $\xi>{1\over 4}$}

Interestingly, as we shall now prove, one can use the characteristic
asymptotic behavior (\ref{Eq16}) of the radial scalar eigenfunction
in order to exclude the existence of asymptotically flat static
matter configurations made of nonminimally coupled massive scalar
fields in the regimes $\xi<0$ and $\xi>1/4$.

Substituting the spatially regular asymptotic eigenfunction
(\ref{Eq16}) of the massive scalar fields into (\ref{Eq9}) and
(\ref{Eq11}), and using the asymptotic spatial behavior (\ref{Eq2})
of the metric components of the asymptotically flat static spacetime
(\ref{Eq1}), one finds the simple functional expressions
\begin{equation}\label{Eq18}
T^{t}_{t}=(4\xi-1)\mu^2\psi^2\cdot[1+O(M/r)]\
\end{equation}
and
\begin{equation}\label{Eq19}
T^{t}_{t}-T^{\phi}_{\phi}=-\xi{{2\mu}\over{r}}\psi^2\cdot[1+O(M/r)]\
\end{equation}
for the components of the energy-momentum tensor. From Eqs.
(\ref{Eq18}) and (\ref{Eq19}) one immediately deduces the far-region
relation
\begin{equation}\label{Eq20}
\text{sgn}(T^{t}_{t})=-\text{sgn}(T^{t}_{t}-T^{\phi}_{\phi})\ \ \ \
\text{for}\ \ \ \ \xi<0 \ \ \text{or}\ \ \xi>1/4\  .
\end{equation}
We point out that the scalar field relation (\ref{Eq20}) {\it
violates} the characteristic relation (\ref{Eq13}) which, as
discussed in \cite{BekMay}, is imposed by causality on the
energy-momentum components of physically acceptable spacetimes. One
therefore concludes that spherically symmetric reflecting stars {\it
cannot} support static massive scalar fields nonminimally coupled to
gravity in the regimes $\xi\leq0$ \cite{Notex0} and $\xi>1/4$.

Before proceeding, we would like to stress the fact that in the
present subsection we have made {\it no} reference to a particular
physical boundary condition at the surface of the central compact
object. Thus, the compact no-massive-scalar-hair theorem presented
in this subsection rules out the existence of spatially regular
static matter configurations made of nonminimally coupled massive
fields in the regimes $\xi<0$ and $\xi>1/4$ for both classical
black-hole spacetimes with absorbing event {\it horizons} and for
{\it horizonless} regular spacetimes describing compact stars with
reflecting surfaces.

\subsection{The no-massive-scalar-hair theorem for spherically symmetric neutral reflecting
stars with positive values of the physical coupling parameter $\xi$}

In the present subsection we shall prove that spherically symmetric
{\it horizonless} compact stars with reflecting boundary conditions
cannot support static matter configurations made of nonminimally
coupled massive scalar fields with generic {\it positive} values of
the physical coupling parameter $\xi$.

Using the Einstein relation $R=-8\pi T$ \cite{Notett}, one obtains
from Eqs. (\ref{Eq9}), (\ref{Eq10}), and (\ref{Eq11}) the functional
expression
\begin{equation}\label{Eq21}
R=-{{8\pi}\over{1-8\pi\xi\psi^2}}\Big\{e^{-\lambda}\Big[\xi\big({{12}\over{r}}+3\nu{'}-3\lambda{'}\big)\psi\psi{'}+
6\xi\psi\psi{''}+(6\xi-1)(\psi{'})^2\Big]-2\mu^2\psi^2\Big\}\
\end{equation}
for the Ricci scalar curvature which characterizes the static curved
spacetime (\ref{Eq1}). Substituting the radial functional expression
(\ref{Eq21}) into the characteristic differential equation
(\ref{Eq6}) of the nonminimally coupled massive scalar fields, one
finds the radial scalar equation
\begin{equation}\label{Eq22}
\psi{''}\cdot\big[1+8\pi\xi(6\xi-1)\psi^2\big]
+\psi{'}\cdot\Big[{1\over2}\big({{4}\over{r}}+\nu{'}-\lambda{'}\big)\big[1+8\pi\xi(6\xi-1)\psi^2\big]
+8\pi\xi(6\xi-1)\psi\psi{'}\Big]
-\mu^2e^{\lambda}\big(1+8\pi\xi\psi^2\big)\psi=0\ .
\end{equation}
%\begin{equation}\label{Eq16}
%\psi{''}\cdot{{1+8\pi\xi(6\xi-1)\psi^2}\over{1-8\pi\xi\psi^2}}
%+\psi{'}\cdot\Big[{1\over2}\big({{4}\over{r}}+\nu{'}-\lambda{'}\big){{1+8\pi\xi(6\xi-1)\psi^2}\over{1-8\pi\xi\psi^2}}
%+{{8\pi\xi(6\xi-1)\psi\psi{'}} \over{1-8\pi\xi\psi^2}}\Big]
%-\mu^2e^{\lambda}{{1+8\pi\xi\psi^2}\over{1-8\pi\xi\psi^2}}\psi=0\ .
%\end{equation}
%\begin{equation}\label{Eq16}
%\psi{''}\cdot{{1+8\pi\xi(6\xi-1)\psi^2}\over{1-8\pi\xi\psi^2}}
%+\psi{'}\cdot\Big\{{1\over2}\big({{4}\over{r}}+\nu{'}-\lambda{'}\big)+{{8\pi\xi\psi}
%\over{1-8\pi\xi\psi^2}}\Big[\xi\big({{12}\over{r}}+3\nu{'}-3\lambda{'}\big)\psi+(6\xi-1)\psi{'}\Big]\Big\}
%-\mu^2e^{\lambda}{{1+8\pi\xi\psi^2}\over{1-8\pi\xi\psi^2}}\psi=0\ .
%\end{equation}
%\begin{equation}\label{Eq16}
%\psi{''}\cdot\Big(1+{{48\pi\xi^2\psi^2}\over{1-8\pi\xi\psi^2}}\Big)
%+\psi{'}\cdot\Big\{{1\over2}\big({{4}\over{r}}+\nu{'}-\lambda{'}\big)+{{8\pi\xi\psi}
%\over{1-8\pi\xi\psi^2}}\Big[\xi\big({{12}\over{r}}+3\nu{'}-3\lambda{'}\big)\psi+(6\xi-1)\psi{'}\Big]\Big\}
%-\mu^2\Big(1+{{16\pi\xi\psi^2}\over{1-8\pi\xi\psi^2}}\Big)e^{\lambda}\psi=0\
%.
%\end{equation}
The (rather cumbersome) non-linear differential equation
(\ref{Eq22}) determines the spatial behavior of the static
nonminimally coupled massive scalar fields in the spherically
symmetric curved spacetime.

We shall now prove that the radial function
\begin{equation}\label{Eq23}
{\cal F}(r;\xi)\equiv 1+8\pi\xi(6\xi-1)\psi^2\
\end{equation}
that appears on the l.h.s of (\ref{Eq22}) is positive definite. This
is obviously true in the physical regimes $\xi\geq 1/6$ and
$\xi\leq0$. As we shall now show explicitly, the physical regime
$0<\xi<1/6$ is also characterized by the inequality ${\cal F}>0$. We
first note that the radial function (\ref{Eq23}) is characterized by
the simple relation [see Eq. (\ref{Eq7})]
\begin{equation}\label{Eq24}
{\cal F}(r=R_{\text{s}})=1\
\end{equation}
at the reflecting surface of the horizonless compact star. Now, let
us assume that ${\cal F}(r)$ vanishes at some point $r=r_0$. Then,
from Eq. (\ref{Eq22}) one finds the simple functional relation
\begin{equation}\label{Eq25}
8\pi\xi(6\xi-1)(\psi{'})^2=\mu^2e^{\lambda}(1+8\pi\xi\psi^2)\ \ \ \
\text{at}\ \ \ \ r=r_0\
\end{equation}
at the assumed root of ${\cal F}(r)$. A simple inspection of Eq.
(\ref{Eq25}) reveals that, in the physical regime $0<\xi<1/6$, the
functional expression on the l.h.s of (\ref{Eq25}) is non-positive
whereas the functional expression on the r.h.s of (\ref{Eq25}) is
positive definite. One therefore concludes that the radial function
${\cal F}(r)$ {\it cannot} switch signs. In particular, taking
cognizance of the boundary relation (\ref{Eq24}), we find the
characteristic inequality
\begin{equation}\label{Eq26}
{\cal F}(r)>0\  .
\end{equation}

Interestingly, from Eqs. (\ref{Eq17}), (\ref{Eq22}), and
(\ref{Eq23}) one finds the compact functional relation
\begin{equation}\label{Eq27}
{\cal F}\cdot\psi\psi{''}=\mu^2e^{\lambda}(1+8\pi\xi\psi^2)\psi^2\ \
\ \ \text{at}\ \ \ \ r=r_{\text{peak}}\
\end{equation}
at the extremum point $r=r_{\text{peak}}$ of the scalar
eigenfunction (where $\psi{'}=0$). Taking cognizance of the
analytically derived inequality (\ref{Eq26}) and the characteristic
relation $\psi\psi{''}<0$ at the extremum point of the static
nonminimally coupled massive scalar field configuration [see Eq.
(\ref{Eq17})], one finds that the functional expression on the l.h.s
of (\ref{Eq27}) is negative definite whereas, for $\xi\geq0$, the
functional expression on the r.h.s of (\ref{Eq27}) is positive
definite. Thus, the characteristic differential relation
(\ref{Eq22}) for the nonminimally coupled massive scalar fields {\it
cannot} be respected at the extremum point $r=r_{\text{peak}}$ of
the scalar eigenfunction. We therefore conclude that spherically
symmetric reflecting stars {\it cannot} support spatially regular
static configurations made of massive scalar fields nonminimally
coupled to gravity in the physical regime $\xi\geq0$ \cite{NoteNew}.

\section{Summary}

The physically important and mathematically elegant no-hair theorems
of Bekenstein and Mayo \cite{BekMay,Bek20} have revealed the
interesting fact that spherically symmetric black holes with
absorbing horizons cannot support asymptotically flat non-linear
matter configurations made of massive scalar fields \cite{Notesm}
nonminimally coupled to gravity in the physical regimes $\xi<0$ and
$\xi\geq {1\over2}$.

Intriguingly, it has recently been revealed that asymptotically flat
{\it horizonless} spacetimes may share the no-hair property with the
more familiar absorbing black-hole spacetimes
\cite{Hodns,Bha,Hodpd}. In particular, the no-hair theorems
presented in \cite{Hodns,Bha,Hodpd} have explicitly proved that
regular curved spacetimes describing compact stars with repulsive
({\it reflecting}) boundary conditions [as opposed to the attractive
({\it ingoing}) boundary conditions which characterize the horizons
of the classical black-hole spacetimes considered in the original
no-hair theorems \cite{BekMay,Bek20}] cannot support non-linear
matter configurations made of minimally coupled scalar, vector, and
tensor fields \cite{Hodns,Bha}, as well as nonminimally coupled
massless scalar fields \cite{Hodpd}.

It is certainly of physical interest to explore the regime of
validity of the no-hair behavior recently observed for regular
horizonless compact objects \cite{Hodns,Bha,Hodpd}. In the present
paper, by studying analytically the non-linear Einstein-scalar field
equations, we have explicitly proved that horizonless neutral stars
with reflecting (that is, {\it repulsive} rather than {\it
absorbing}) boundary conditions cannot support spatially regular
non-linear matter configurations made of {\it massive} scalar fields
{\it nonminimally} coupled to gravity \cite{Notebu}.

Finally, we would like to emphasize the interesting fact that, while
existing no-hair theorems for the composed
black-hole-massive-scalar-field system are valid in the restricted
physical regimes $\xi<0$ and $\xi\geq1/2$
\cite{BekMay,Bek20,Notenv,Notesm}, the no-hair theorem presented in
this paper for the composed
compact-reflecting-star-massive-scalar-field system is valid for
{\it generic} values of the dimensionless nonminimal coupling
parameter $\xi$.

%\newpage

\bigskip
\noindent
{\bf ACKNOWLEDGMENTS}
\bigskip

This research is supported by the Carmel Science Foundation. I would
like to thank Yael Oren, Arbel M. Ongo, Ayelet B. Lata, and Alona B.
Tea for helpful discussions.

%\newpage


\begin{thebibliography}{99}

\bibitem{Bek1} J. D. Bekenstein, Phys. Rev. Letters {\bf 28}, 452 (1972).

\bibitem{Teit} C. Teitelboim, Lett. Nuov. Cim. {\bf 3}, 326 (1972).

\bibitem{Chas} J. E. Chase, Commun. Math. Phys. {\bf 19}, 276 (1970).

\bibitem{Bekto} J. D. Bekenstein, Physics Today {\bf 33}, 24 (1980).

\bibitem{Heu} M. Heusler, J. Math. Phys. {\bf 33}, 3497 (1992);
M. Heusler, Class. Quant. Grav. {\bf 12}, 779 (1995).

\bibitem{Bek2} J. D. Bekenstein, Phys. Rev. D {\bf 51}, R6608 (1995).

\bibitem{BekMay} A. E. Mayo and J. D. Bekenstein, Phys. Rev. D {\bf 54}, 5059 (1996).

\bibitem{Bek20} J. D. Bekenstein, arXiv:gr-qc/9605059 .

\bibitem{Noteev} It is worth noting that the no-scalar-hair property
deduced in \cite{Bek1,Teit,Bekto,Chas,Heu,Bek2,BekMay,Bek20} can be
extended to the physically interesting regime of non-linear external
matter configurations with monotonically increasing self-interaction
scalar potentials \cite{Bek20,BekMay}.

\bibitem{Noteit} It is worth mentioning that, while the elegant and
physically important no-hair theorems of \cite{Bek1} and \cite{Teit}
have ruled out the existence of hairy black-hole spacetimes
supporting external static configurations made of massive scalar
fields, it has been explicitly proved in \cite{Hodrc,Herkr} that
spinning black-hole spacetimes can support spatially regular {\it
stationary} (that is, non-decaying in time) massive scalar (bosonic)
fields.

\bibitem{Hodrc} S. Hod, Phys. Rev. D {\bf 86}, 104026 (2012) [arXiv:1211.3202];
S. Hod, The Euro. Phys. Journal C {\bf 73}, 2378 (2013)
[arXiv:1311.5298]; S. Hod, Phys. Rev. D {\bf 90}, 024051 (2014)
[arXiv:1406.1179]; S. Hod, Phys. Lett. B {\bf 739}, 196 (2014)
[arXiv:1411.2609]; S. Hod, Class. and Quant. Grav. {\bf 32}, 134002
(2015) [arXiv:1607.00003]; S. Hod, Phys. Lett. B {\bf 751}, 177
(2015); S. Hod, Class. and Quant. Grav. {\bf 33}, 114001 (2016); S.
Hod, Phys. Lett. B {\bf 758}, 181 (2016) [arXiv:1606.02306]; S. Hod
and O. Hod, Phys. Rev. D {\bf 81}, 061502 Rapid communication (2010)
[arXiv:0910.0734]; S. Hod, Phys. Lett. B {\bf 708}, 320 (2012)
[arXiv:1205.1872]; S. Hod, Jour. of High Energy Phys. {\bf 01}, 030
(2017) [arXiv:1612.00014].

\bibitem{Herkr} C. A. R. Herdeiro and E. Radu, Phys. Rev. Lett. {\bf 112}, 221101
(2014); C. L. Benone, L. C. B. Crispino, C. Herdeiro, and E. Radu,
Phys. Rev. D {\bf 90}, 104024 (2014); C. A. R. Herdeiro and E. Radu,
Phys. Rev. D {\bf 89}, 124018 (2014); C. A. R. Herdeiro and E. Radu,
Int. J. Mod. Phys. D {\bf 23}, 1442014 (2014); Y. Brihaye, C.
Herdeiro, and E. Radu, Phys. Lett. B {\bf 739}, 1 (2014); J. C.
Degollado and C. A. R. Herdeiro, Phys. Rev. D {\bf 90}, 065019
(2014); C. Herdeiro, E. Radu, and H. R\'unarsson, Phys. Lett. B {\bf
739}, 302 (2014); C. Herdeiro and E. Radu, Class. Quantum Grav. {\bf
32} 144001 (2015); C. A. R. Herdeiro and E. Radu, Int. J. Mod. Phys.
D {\bf 24}, 1542014 (2015); C. A. R. Herdeiro and E. Radu, Int. J.
Mod. Phys. D {\bf 24}, 1544022 (2015); P. V. P. Cunha, C. A. R.
Herdeiro, E. Radu, and H. F. R\'unarsson, Phys. Rev. Lett. {\bf
115}, 211102 (2015); B. Kleihaus, J. Kunz, and S. Yazadjiev, Phys.
Lett. B {\bf 744}, 406 (2015); C. A. R. Herdeiro, E. Radu, and H. F.
R\'unarsson, Phys. Rev. D {\bf 92}, 084059 (2015); C. Herdeiro, J.
Kunz, E. Radu, and B. Subagyo, Phys. Lett. B {\bf 748}, 30 (2015);
C. A. R. Herdeiro, E. Radu, and H. F. R\'unarsson, Class. Quant.
Grav. {\bf 33}, 154001 (2016); C. A. R. Herdeiro, E. Radu, and H. F.
R\'unarsson, Int. J. Mod. Phys. D {\bf 25}, 1641014 (2016); Y.
Brihaye, C. Herdeiro, and E. Radu, Phys. Lett. B {\bf 760}, 279
(2016); Y. Ni, M. Zhou, A. C. Avendano, C. Bambi, C. A. R. Herdeiro,
and E. Radu, JCAP {\bf 1607}, 049 (2016); M. Wang, arXiv:1606.00811
.

\bibitem{Notewx} Here $\xi$ is a dimensionless physical
parameter which quantifies the strength of the nonminimal coupling
of the field to the scalar curvature of the spacetime [see Eq.
(\ref{Eq3}) below].

\bibitem{Notenv} It is worth stressing the fact that, to the best of our knowledge, at present
there is no mathematically rigorous no-hair theorem which rules out
the possible existence of hairy black-hole spacetimes supporting
static matter configurations made of neutral nonminimally coupled
{\it massive} scalar fields in the physical regime $0<\xi<1/2$
\cite{BekMay}.

\bibitem{NoteBB} In this context, it is important to mention the existence of the extremal
Bocharova-Bronnikov-Melnikov-Bekenstein (BBMB) \cite{BB,BekenBB}
hairy black-hole solution of the non-linearly coupled
Einstein-scalar field equations. This spherically symmetric hairy
black-hole spacetime is characterized by a conformally coupled
massless scalar field with $\xi=1/6$. Obviously, this value of the
dimensionless nonminimal coupling parameter $\xi$ is outside the
regimes $\xi<0$ and $\xi\geq1/2$ covered by the influential no-hair
theorems presented in \cite{BekMay,Bek20} for nonminimally coupled
scalar fields.

\bibitem{BB} N. Bocharova, K. Bronikov and V. Melnikov, Vestn. Mosk. Univ. Fiz. Astron. {\bf 6}, 706 (1970).

\bibitem{BekenBB} J. D. Bekenstein, Ann. Phys. (NY) {\bf 82}, 535 (1974);
J. D. Bekenstein, Ann. Phys. (NY) {\bf 91}, 72 (1975).


\bibitem{Notesm} It is worth noting that the no-hair theorems
presented in \cite{BekMay,Bek20} are valid for scalar fields with
positive semidefinite self-interaction potentials.

\bibitem{Hodns} S. Hod, Phys. Rev. D {\bf 94}, 104073 (2016) [arXiv:1612.04823].

\bibitem{MPF} E. Maggio, P. Pani, and V. Ferrari, arXiv:1703.03696.

\bibitem{BCP} R. Brito, V. Cardoso, and P. Pani, Lect. Notes Phys. 906, 1 (2015).

\bibitem{Hodrj} S. Hod, Jour. of High Energy Phys. {\bf 06}, 132 (2017)
[arXiv:1704.05856].

\bibitem{Hodrb} S. Hod, Phys. Lett. B {\bf 770}, 186 (2017).

\bibitem{Noteref} The term `reflecting star' is used here to describe
a physical compact object for which the external matter fields
vanish on its outer reflecting surface [see Eq. (\ref{Eq7}) below].

\bibitem{Bha} S. Bhattacharjee and S. Sarkar, Phys. Rev. D {\bf 95},
084027 (2017).

\bibitem{Hodpd} S. Hod, Phys. Rev. D {\bf 96}, 024019 (2017).

\bibitem{Notecor} Here we use the Schwarzschild spacetime coordinates $(t,r,\theta,\phi)$.

\bibitem{Noteunit} We shall use natural units in which $G=c=1$.

\bibitem{NoteSEH} Here $S_{\text{EH}}$ is the Einstein-Hilbert action.

\bibitem{Notemu} Note that the mass parameter $\mu$, which characterizes the
nonminimally coupled massive scalar field, stands for $\mu/\hbar$.
Thus, this physical parameter has the dimensions of (length)$^{-1}$.

\bibitem{Notetag} Here a prime ${'}$ denotes a spatial derivative with respect to $r$.

\bibitem{Noterefbs} It is worth mentioning that, following the influential work of Press and
Teukolsky \cite{PressTeu2} on the `black-hole bomb' phenomenon, many
researches have explored the physical properties of the composed
black-hole-scalar-field-reflecting-mirror system. In this composed
physical system, one places a reflecting surface around the black
hole whose role is to prevent the scalar field from radiating its
energy to infinity. On the other hand, in the present paper (see
also \cite{Hodns,Bha,Hodpd}) we consider a spherically symmetric
reflecting surface whose role is to prevent the nonminimally coupled
massive scalar field from radiating its energy into the central
horizonless compact star.

\bibitem{PressTeu2} W. H. Press and S. A. Teukolsky, Nature {\bf
238}, 211 (1972); W. H. Press and S. A. Teukolsky, Astrophys. J.
{\bf 185}, 649 (1973).

\bibitem{Notex0} The no-hair theorem of \cite{Hodns} can be used to rule out the
existence of external static matter configurations made of minimally
coupled ($\xi=0$) massive scalar fields.

\bibitem{Notett} Here $T$ is the trace of the energy-momentum tensor.

\bibitem{NoteNew} Thus far we have considered horizonless reflecting stars
with Dirichlet [that is, $\psi(r=R_{\text{s}})=0$] boundary
conditions. It is interesting to note that one can extend the regime
of validity of the no-hair behavior to the case of compact stars
with Neumann [that is, $\psi{'}(r=R_{\text{s}})=0$] boundary
conditions. In particular, as emphasized above, the no-hair theorem
presented in section IIIA is valid in the regimes $\xi<0$ and
$\xi>1/4$ for {\it generic} inner boundary conditions. In addition,
one finds from Eq. (\ref{Eq22}) that, for the case of Neumann
boundary conditions, the functional relation (\ref{Eq27}) is valid
at the surface $r=R_{\text{s}}$ of the reflecting star. Furthermore,
using the fact that ${\cal F}(r)$ cannot switch signs together with
the asymptotic behavior ${\cal F}(r/M\to\infty)\to1$ [see Eqs.
(\ref{Eq16}) and (\ref{Eq23})], one concludes that ${\cal F}(r)>0$
for the Neumann case as well [that is, the characteristic inequality
(\ref{Eq26}) is valid for both Dirichlet and Neumann boundary
conditions]. Now, if $\psi\psi{''}>0$ at the surface of the star,
then from the characteristic asymptotic behavior (\ref{Eq16}) of the
nonminimally coupled massive scalar field one learns that the radial
eigenfunction $\psi(r)$ must have (at least) one extremum point
$r_{\text{peak}}\in(R_{\text{s}},\infty)$ with the functional
properties (\ref{Eq17}). At this extremum point, the l.h.s of
(\ref{Eq27}) is negative definite whereas, for $\xi\geq0$, the r.h.s
of (\ref{Eq27}) is positive definite. Thus, one concludes that the
characteristic relation (\ref{Eq22}) {\it cannot} be respected at
the extremum point $r=r_{\text{peak}}$ of the scalar eigenfunction.
Likewise, if $\psi\psi{''}<0$ at the boundary $r=R_{\text{s}}$ of
the star (where $\psi{'}=0$ for the Neumann boundary conditions),
then one finds that, on the reflecting surface, the l.h.s of
(\ref{Eq27}) is negative definite whereas, for $\xi\geq0$, the r.h.s
of (\ref{Eq27}) is positive definite. Again, one realizes that the
characteristic relation (\ref{Eq22}) {\it cannot} be respected at
the compact reflecting surface $r=R_{\text{s}}$ of the star. We
therefore conclude that spherically symmetric horizonless reflecting
stars with Neumann boundary conditions {\it cannot} support static
configurations made of massive scalar fields nonminimally coupled to
gravity.

\bibitem{Notebu} It is worth stressing the fact that the no-hair
theorem presented in section IIIA for massive scalar fields
nonminimally coupled to gravity in the physical regimes $\xi<0$ and
$\xi>{1\over 4}$ is valid for {\it generic} inner boundary
conditions. In particular, this theorem rules out the existence of
massive scalar hair with nonminimal coupling parameter in the
dimensionless regimes $\xi<0$ and $\xi>1/4$ for both horizonless
curved spacetimes describing compact regular stars and for
black-hole spacetimes with absorbing event horizons.

\end{thebibliography}
\end{document}